# A distributed data warehouse system for astroparticle physics

**Minh-Duc Nguyen[1, a], Alexander Kryukov[1], Julia Dubenskaya[1], Elena Korosteleva[1], Stanislav Polyakov[1], Evgeny Postnikov[1], Igor Bychkov[2], Andrey Mikhailov[2], Alexey Shigarov[2], Oleg Fedorov[3], Yulia Kazarina[3], Dmitry Shipilov[3], Dmitry Zhurov[3]**

[1] *Lomonosov Moscow State University, Skobeltsyn Institute of Nuclear Physics, Moscow, Russia*

[2] *Matrosov Institute for System Dynamics and Control Theory, Siberian Branch of Russian Academy of Sciences, Irkutsk, Russia*

[3] *Applied Physics Institute, Irkutsk State University, Irkutsk, Russia*

E-mail: [a] nguyendmitri@gmail.com

A distributed data warehouse system is one of the actual issues in the field of astroparticle physics. Famous experiments, such as TAIGA, KASCADE-Grande, produce tens of terabytes of data measured by their instruments. It is critical to have a smart data warehouse system on-site to store the collected data for further distribution effectively. It is also vital to provide scientists with a handy and user-friendly interface to access the collected data with proper permissions not only on-site but also online. The latter case is handy when scientists need to combine data from different experiments for analysis. In this work, we describe an approach to implementing a distributed data warehouse system that allows scientists to acquire just the necessary data from different experiments via the Internet on demand. The implementation is based on CernVM-FS with additional components developed by us to search through the whole available data sets and deliver their subsets to users' computers.

# 1. Introduction

In this paper, we describe the current status of the data warehouse system at Astroparticle.online. Astroparticle.online [1] is a project funded by the German-Russian Astroparticle Data Life Cycle Initiative to support physical experiments in astroparticle physics such as TAIGA [2] and KASCADE-Grande [3]. The primary goal of the project is to create a smart mechanism to distribute data at each site where the experiments take place to scientists so that they can use together data from different experiments in their research for analysis as if the files are locally available in their computers.

There are some requirements for such a data warehouse system due to the computing facility at the experiments' sites. First of all, the implementation of the system must not lead to any changes in the existing hardware and software infrastructure at the sites. All new components must be added to the sites as independent modular add-ons. Second, all work that might cause high CPU loading to the computing facility of the cites should be done somewhere else. Third, only the exact amount of necessary data should be transferred from the sites due to limited bandwidth and a large amount of accumulated data. And finally, the data are read-only for all users. Changes to data must be done only on-site not online.

All described above requirements narrow the searching scope and lead to three major existing open-source solutions: CernVM-FS [4], HDFS [5] and OpenAFS [6]. In this paper, we will describe our attempt to build the targeted data warehouse system using CernVM-FS and the problems we are facing during the process. The structure of this paper is as follow. In the second section, we take a deep dive into CernVM-FS to see how it works. In the third section, we explain how we build our data warehouse system using CernVM-FS. In the last section, we point out what we managed to do and the future plan.

# 2. CernVM-FS

CernVM-FS is a widely used file system at CERN, the European Organization for Nuclear Research, to distribute the software to be used in physical experiments to scientists. The key idea is that CernVM-FS indexes only the metadata and the directory structure and distribute them to users. The metadata and directory structure are very small, so it is easier and faster to be transferred. Users can browse through the remote catalogue right after mounting it to a mount point in the OS using CernVM-FS client. The content of a file is fetched and delivered to users only on actual reads so a lot of unnecessary data transfer can be avoided.

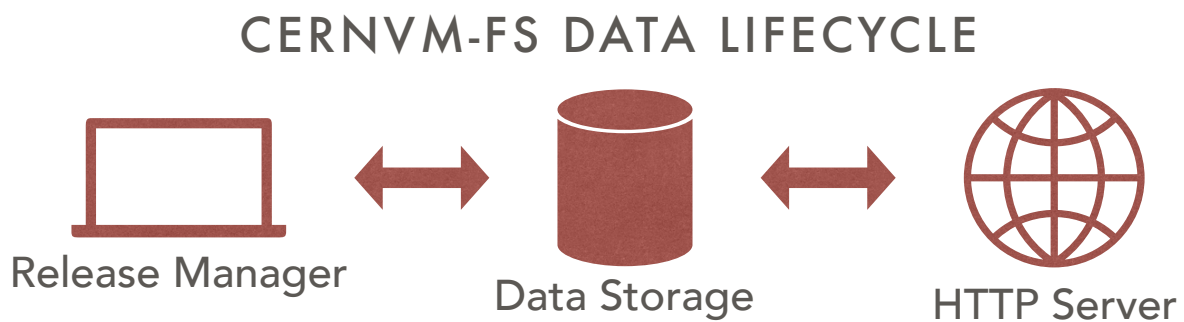

Figure 1. CernVM-FS data life cycle scheme.

The data life cycle in Cern VM-FS is one directional. Data are available as read-only to users. Changes can be made only from the server side. After a careful review, a data release manager creates a new release of the data and upload it to the central repository controlled by a CernVM-FS server. This process is called a transaction. During a transaction, the new catalogue of files and folders is indexed and added as flat objects into a "big bag" similar to a Git-repository [7]. After the confirmation from the release manager, the transaction is published.

## CERNVM-FS DATA DISTRIBUTION

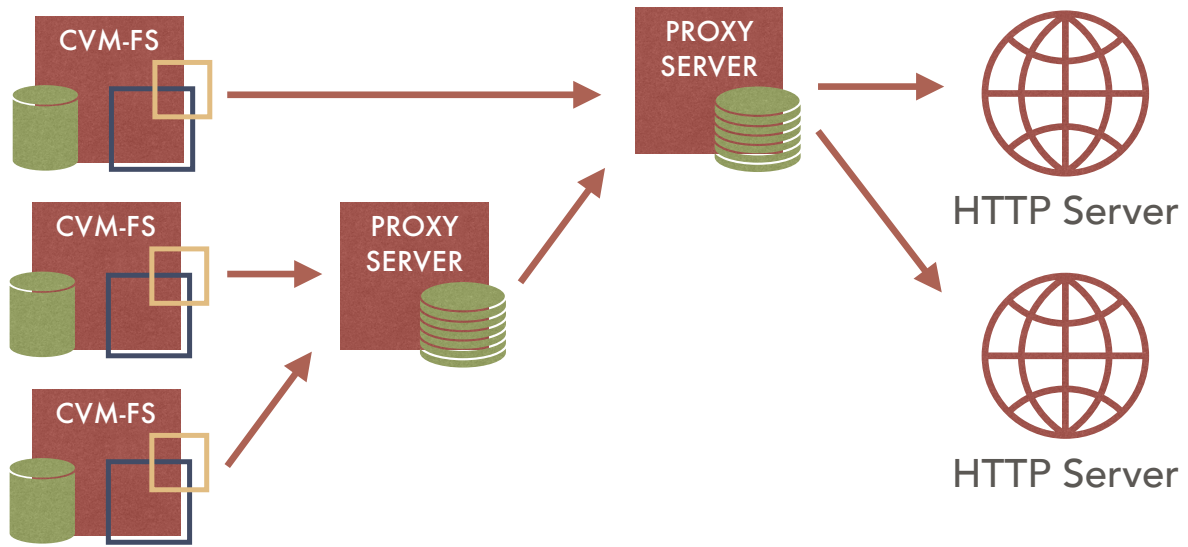

Figure 2. CernVM-FS data distribution scheme.

The CernVM-FS repository is distributed through an https server. In terms of CernVM-FS an https server at the root level where the data catalogue itself is located is called stratrum-0. Since the content is distributed through an https server, it can be cached and redistributed through a proxy server. The redistribution proxy server is called stratum-1. Different combinations of stratum-0 and stratum-1 servers are possible as shown in figure 2. A user can then use the CernVM-FS client to mount the remote catalogue to a mount point in her computer. The content of files is fetched from the stratum-0 or stratum-1 server to the user's machine on-read. As long as the files are in the local cache of the CernVM-FS client they can be accessed offline.

## 3. The astroparticle.online data warehouse system

At each site of the physical experiments (TAIGA and KASCADE-Grande), the result after each run is saved as files in specific binary formats. When a physicist investigates an event, she collects all files related to the event and analyses them. The process when the physicist searches for necessary files is called preliminary search. The goal of the data warehouse system is to make this process as fast as possible. Once necessary files have been found, we would want the files to be able to be reused as long as possible without retransferring them from the root server. We also want to implement a light data access policy to prevent users from seeing the entire catalogue of data from all sites. Users can see only files related to their queries each time. Taking into account all requirements above, we started building a prototype of the targeted data warehouse system based on CernVM-FS. The overall architecture of the system is shown below.

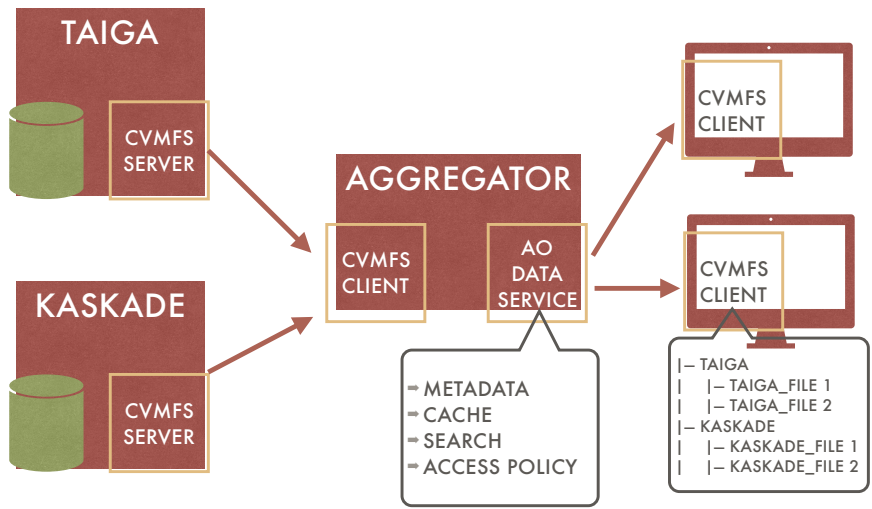

Figure 3. The astroparticle.online data warehouse system based on CernVM-FS

Each data storage server at each experiment site contains an instance of the CernVM-FS server. The existing data storage is plugged into Cern VM-Fs server as an external data source. In this case, CernVM-FS is still able to index the data catalogue, but the data catalogue itself is still managed by the old accepted way at the site. We don't want to make a full copy of the whole data catalogue in another partition for indexing it by CernVM-FS. All stratum-0 instance data catalogues are mounted to a stratum-1 proxy server. We call it the aggregator. At this level, we fetch the metadata from the raw files and create the indices based on a set of parameters that later are used by users in their search queries such as event start time, event end time, energy range, site location etc. The data access policy is also implemented at the aggregator level. Another critical function of the aggregator is caching the most frequently accessed files and search result. The aggregator also serves as an API server. From the user side, one can use a customised client that talks with the aggregator using the API to find a collation of files, then the client uses CernVM-FS client to mount those files to the local computer.

## 4. Conclusion

At the moment we were able to merge all data catalogues from different sites into a big one and redistribute it to users. Data caching is done by CernVM-FS out-of-the-box. Metadata indexing is still an ongoing work. The most significant issue with CernVM-FS is that the whole data catalogue is visible and accessible for everyone by design. We are looking for a solution to implement the data access policy that requires a minimum change in the CernVM-FS core.

In the future, we plan to build other prototypes using HDFS and OpenAFS with the same requirements. After that, we will compare the prototypes by metrics including the complexity, spent time and effort to adapt each solution to build the target data warehouse system and to maintain it. We also plan to conduct a performance benchmark.

## Acknowledgment

This work is supported by the RSF grant #18-41-06003.